\documentclass{article} 
\usepackage{iclr2021_conference,times}

\usepackage{amsmath,amsfonts,bm}

\def\eqref#1{equation~\ref{#1}}

\def\1{\bm{1}}

\DeclareMathAlphabet{\mathsfit}{\encodingdefault}{\sfdefault}{m}{sl}
\SetMathAlphabet{\mathsfit}{bold}{\encodingdefault}{\sfdefault}{bx}{n}

\usepackage{hyperref}
\usepackage{url}
\usepackage{comment}
\usepackage{graphicx}
\usepackage{wrapfig}
\usepackage{graphicx,subcaption}
\usepackage{hyperref}
\usepackage{siunitx}  
\iclrfinalcopy

\title{Invertible Surrogate Models: Joint surrogate modelling and reconstruction of Laser-Wakefield Acceleration by invertible neural networks}

\author{Friedrich Bethke$^1$, Richard Pausch$^1$, Patrick Stiller$^1$, \\ \textbf{Alexander Debus}$^1$, \textbf{Michael Bussmann}$^{2,1}$, \textbf{Nico Hoffmann}$^1$\\
\\
$^1$ Helmholtz-Zentrum Dresden - Rossendorf, Bautzner Landstrasse 400, 01328 Dresden, Germany\\
$^2$ CASUS - Center for Advanced Systems Understanding, Untermarkt 20, 02826 Görlitz, Germany\\
\texttt{\{f.bethke, n.hoffmann\}@hzdr.de} 
}
\begin{document}

\maketitle

\begin{abstract}
Invertible neural networks are a recent technique in machine learning promising neural network architectures that can be run in forward and reverse mode. In this paper, we will be introducing invertible surrogate models that approximate complex forward simulation of the physics involved in laser plasma accelerators: iLWFA. The bijective design of the surrogate model also provides all means for reconstruction of experimentally acquired diagnostics. The quality of our invertible laser wakefield acceleration network will be verified on a large set of numerical LWFA simulations.
\end{abstract}


\section{Introduction}
Laser plasma accelerators are a new technique to accelerate charged particles, providing acceleration gradients orders of magnitude higher than with conventional accelerators. 
One of the most established techniques is the so-called laser wakefield acceleration (LWFA) \citep{Tajima1979,Esarey2009}.
A short and intense laser pulse drives a plasma wave inside which electrons can be accelerated. 
Especially in the so-called blowout or bubble regime \citep{Pukhov2002,Geddes2004,Faure2004,Mangles2004}, quasi-monoenergetic electron bunches with pulse duration and extent in the range of a few \SI{}{\femto \second} and few \SI{}{\micro \meter} can be generated which are not achievable by conventional accelerators.
One of the most commonly used methods to inject electrons into the plasma accelerator is downramp injection \citep{Schmid2010,Buck2013,Swanson2017}, during which the plasma wake transitions from a high- to a low-density region, thereby expanding the plasma cavity and, if the laser is strong enough, injecting electrons that are further accelerated.
This injection method can provide significantly better beam quality compared to other injection methods \citep{Barber2017}. 
The injection process depends both on the laser evolution as well as on the plasma dynamics, which are nonlinearly coupled.
During the consecutive acceleration, the beam quality is influenced by the complex interplay between beam and wake, called beam loading \cite{Couperus2017}.
Making predictions for the injection process and acceleration thus require modeling with large-scale particle-in-cell simulations \citep{MartinezdelaOssa2017}. 
A surrogate model for these simulations is highly desireable since the simulations are highly computationally demanding even for highly-optimized particle-in-cell codes\cite{Burau2010}, A fast prediction of electron parameters of interest could accelerate the applicability of down-ramp injection methods even further.  
Main contributions of this paper is the introduction of invertible surrogate modelling applied to accelerator physics. The proposed invertible LWFA surrogate model (iLWFA) approximates the non-linear forward simulation of LWFA with focus on down-ramp injection. Furthermore, the choice of architecture allows us to jointly solve the inverse problem, i.e. reconstruction of simulation parameters based on energy spectra.  

\section{Related works}
\begin{figure}[h]
\vskip 0.05in
\begin{center}
\centerline{\includegraphics[width=0.75\columnwidth]{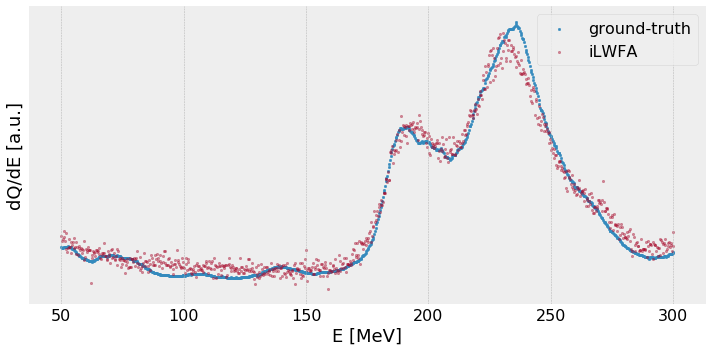}}
\caption{The prediction of the iLWFA network (red) on validation data very closely approximates the energy spectrum of numerical particle-in-cell code (blue) in a fraction of time. The spectrum shows the final charge of the electrons per unit energy $\mathrm{d}Q/\mathrm{d}E \hat{=} \Psi$ over energy $E$. For this particular parameter set $p$, the most electrons end up to have an energy between \SI{180}{\mega \electronvolt} and \SI{250}{\mega \electronvolt}.}
\label{fig:energySpectra}
\end{center}
\vskip -0.05in
\end{figure} 
Computationally expensive simulations are necessary for simulation of complex physical processes. Scanning the input parameters of physical simulations significantly increases the runtime of the simulations, since the simulation has to be cold-started over and over again. Surrogate models which are able to learn the relationship among the input parameters and the simulation output could significantly lower the computational complexity and enables more extensive analysis of the underlying physical system without restarting the simulation every time. 

Neural networks are a valuable choice as surrogate models due to their ability as universal function approximators \citep{HORNIK}. Certain established approaches rely on the multilayer perceptron architecture \citep{pinn,DGM,stiller2020} while more sophisticated approaches such as generative adversial networks, autoencoders \citep{xie2018tempoGAN,deepfluid} and graph neural networks are also used \citep{gnn}. However, all these approaches are designed as surrogates for the forward process of the underlying simulation, so that the inverse mapping must be represented by a separate model (e.g. \citep{hpm}, \citep{deephpm}).

Invertible neural networks are a promising candidate as surrogate models through there ability to learn the forward and inverse mapping simultaneously. \cite{INNsurrogate} applied a conditional invertible neural network as a surrgogate model for the estimation of a non-Gaussian permeability field in multiphase flows. In this paper, we will be learning an invertible conditional mapping between simulation parameters and -outcomes by adapting the architecture of \citep{Ardizzone2018}, which is more challenging to train but promises a more accurate approximation of the forward pass (simulation) and the inverse process (reconstruction).
\section{Method}
A surrogate model of laser wakefield acceleration can be seen as mapping $g(p) = \Psi(E)$ from $n_p$ simulation parameters $p \in \mathbb{R}^{n_p}$ to energy spectrum $\frac{\mathrm{d}Q}{\mathrm{d}E} \hat{=} \Psi =  [\Psi_1,\Psi_2,...,\Psi_{n_E}]$.
The spectrum $\Psi$ is discretized for $n_E$ energy bins $E_i \in [40; 300]$~\SI{}{\mega \electronvolt} and represents the total charge of all electrons in that specific energy bin $E_i$. 
In experiment, parameters $p$ are typically not directly measurable but provide important insights about the state of the system rendering the need for reconstruction $f(\cdot)$ of parameters $p$ given the energy spectrum $\Psi$, i.e. $f(\Psi) = p$. The reconstruction can be seen as (ill-posed) inverse process of our surrogate model which, however, might not be  injective due to loss of information as the energy spectrum might not represent the full state of the system. This means that certain parameters $p_j,p_k$ might map to the same energy spectrum $\Psi$, i.e. $g(p_j)=g(p_k)=\Psi$. This ambiguity implies that any neural network $n \approx f$ would either pick any of the feasible parameters or return the average of both, depending on the choice of architecture and objective function. We approach this challenge by invertible neural networks \citep{Ardizzone2018} that allow us to recover the full posterior predictive distribution of $p$ given the observed energy spectrum $\Psi$, i.e. $\pi_{\mathcal{P}}(p | \Psi, z) = f(\Psi,z)$. The modes of this posterior predictive distribution $\pi_{\mathcal{P}}(p | \Psi, z)$  correspond to all parameter configurations explaining the observed energy spectrum $\Psi$. 
\begin{figure}[t!]
  \begin{subfigure}[t]{0.3\textwidth}
    \includegraphics[width=4.2cm]{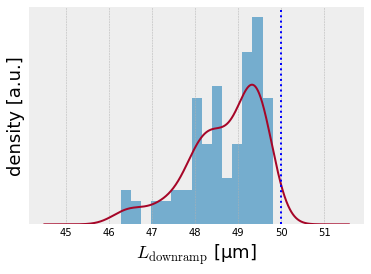}
    \caption{}
  \end{subfigure}
  \hfill
  \begin{subfigure}[t]{0.3\textwidth}
    \includegraphics[width=4.2cm]{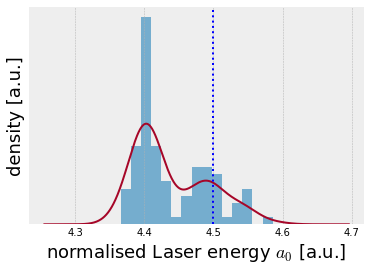}
    \caption{}
  \end{subfigure}
  \hfill
  \begin{subfigure}[t]{0.3\textwidth}
    \includegraphics[width=4.2cm]{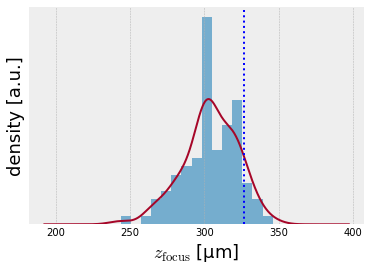}
    \caption{ }
  \end{subfigure}
  
  \caption{The posterior predictive distribution of iLWFA on validation data is close to the ground-truth (blue). The method is not restricted to certain family of distributions as well as number of modes as (a)-(c) show. The two modes of (b) encode information about the ambiguity when reconstructing the laser parameter from a single energy spectrum.}
  \label{fig:parameterReconstruction}
\end{figure}

The main idea of an invertible surrogate model for Laser-Wakefield acceleration is that some  neural networks transform a simple multivariate base distribution $z \sim \pi_\mathcal{Z} = N(0,I)$ with zero mean and identity matrix $I$ into a complex posterior distribution of parameters $\pi_{\mathcal{P}}$ subject to observation $\Psi$ in terms of a non-linear function $f(\Psi, z)$. The posterior can be derived by the change-of-variables approach as of
\begin{equation}
    \label{eqn:NF}
    \pi_\mathcal{P} = \pi_\mathcal{Z} \left| det\left(\frac{\partial f(\Psi,z | \theta)}{\partial [\Psi^T,z^T]}\right) \right|
\end{equation}
given parameters $\theta$ of $f(\cdot)$. This architecture allows us to resolve ambiguous mappings since $f(\cdot)$ is trained such taht it transforms different parts of our base distribution $\pi_\mathcal{Z}$ to each mode of the data distribution while any simpler data-driven approach would either pick one mode or return the average value of all corresponding parameters. It has been shown that this choice of architecture is able to return the true posterior distribution \citep{Ardizzone2018}.

The function $f(\cdot)$ is defined by a composition of $L$ affine coupling transforms \citep{Dinh2016}. The input $u^{(l)}$ to the $l$-th coupling transform is split into two equally sized halves $u^{(l)} = [u_1^{(l)}, u_2^{(l)}]$ and gets transformed into $v^{(l)} = [v_1, v_2]$ according to 
\begin{equation}
\begin{split}
v_1^{(l)} & = u_1^{(l)} \cdot exp(s^{(l)}(u_2^{(l)})) + t^{(l)}(u_2^{(l)}) \\
v_2^{(l)} & = u_2^{(l)} \cdot exp(s^{(l)}(v_1^{(l)})) + t^{(l)}(v_1^{(l)})
\end{split}
\end{equation}
for any block $1 \leq l \leq L$ of coupling transform and fully-connected neural networks $s^{(l)}, t^{(l)}$. This design can be seen as a modified version of the GLOW approach \citep{Kingma2018}.
Finally, we need to introduce some zero-padding to the co-domain of $f: \mathbb{R}^{M_E+M_z} \to \mathbb{R}^{M_p+M_0}$ to learn a bijective mapping, i.e. $[\Psi, z] = u^{(1)} \in \mathbb{R}^{M_E+M_z}$ and $[p,0] = v^{(L)} \in \mathbb{R}^{M_p + M_0}$ with $M_p + M_0 = M_E + M_z$. $M_E$ denotes the number of energy bins of our energy spectra $\Psi$ while $M_p$ is the number of parameters (here: $3$), $M_z$ is the size of our latent variable $z$ and $M_0$ the resulting zero-padding. This choice of coupling transform is easily invertible and leads to a triangular Jacobian structure such that the Jacobian determinant of eqn. \ref{eqn:NF} can be computed computationally inexpensively \citep{Dinh2016}. Additionally, the affine coupling transforms also enable fast inference times of the forward direction (surrogate model), i.e. $g(p) = INN(p,0) = [\Psi, z]$ as well as the reverse direction (reconstruction). 

The bidirectional training of the invertible surrogate model is carried out by minimizing our objective function $L$ with respect to the parameters $\theta$ of all subnetworks $s^{(l)}, t^{(l)}$,  
\begin{equation}
L = L_\Psi + \lambda_p L_p + \lambda_z L_z
\end{equation}
with $L_\Psi$ being the supervised forward loss, $L_p$ is the supervised reconstruction loss and $L_z$ preserves the prior distribution of the latent variable $z$. The Lagrange multipliers $\lambda_p, \lambda_z$ scale each term of our objective function accordingly. The supervised forward loss,
\begin{equation}
L_\Psi = \sum_{i=0}^{N} || \Psi_i - \widehat{\Psi}_i ||_1
\end{equation} enforces similarity of the approximated forward simulation $[\widehat{\Psi}_i, \widehat{z}] = f(p_i, 0)$ with ground-truth energy spectrum $\Psi_i$ of parameter $p_i$. The reconstruction loss $L_p$ reads 
\begin{equation}
L_p = \sum_{i=0}^{N} (||p_i - \widehat{p}_i||_2^2 + ||\widehat{0}_i||_2^2) 
\end{equation}
with $[\widehat{p}_i, \widehat{0}_i] = g(\Psi_i, z)$. All terms are evaluated at at $N$ samples of the training set. The latter term of $L_p$ enforces that the invertible network is not encoding any information into the \textit{predicted} zero-padding $\widehat{0}$ of $g$. Finally, the predicted latent variable $\widehat{z}$ is regularized by maximum mean discrepancy (MMD) as of \citep{Ardizzone2018},
\begin{equation}
L_z = MMD(f(p,0),\pi_\Psi\pi_z)
\end{equation}
to enforce normality of $\widehat{z}$ and independence to the predicted $\widehat{\Psi}$.

\section{Results \& Discussion}
All $2.7$ Terabytes of training data were generated by large scale PIConGPU \citep{Bussmann2013,Burau2010} simulations run by a jupyter-based scheduler \citep{Rudat2019}. 
In the following the 3 input parameters to the PIConGPU simulations are: the laser's normalized field strength $a_0$, the laser's focus position $z_\mathrm{focus}$ with $z_\mathrm{focus} = 105\,\mathrm{\mu m}$  representing a focus position at the density downramp, and the length of the downramp density transition $L_\mathrm{downramp}$ from a density $2\cdot10^{19}\,\mathrm{cm^{-3}}$ to $1.1\cdot10^{19}\,\mathrm{cm^{-3}}$. 
Weights $\lambda$ of our objective function were subject to a hyperparameter optimisation. The best architecture consists of latent distribution $\pi_z$ with $M_z = 20$ and a stack of 10 coupling transforms. Each transform consists of two fully-connected neural networks with four hidden layers amounting to $2.9\, M$ parameters for that transform. The total number of parameter of the invertible surrogate model is $\approx 29\, M$.


\subsection{Surrogate model}
\begin{figure}[!ht]
  \begin{subfigure}[t]{\textwidth}
    \includegraphics[width=12cm]{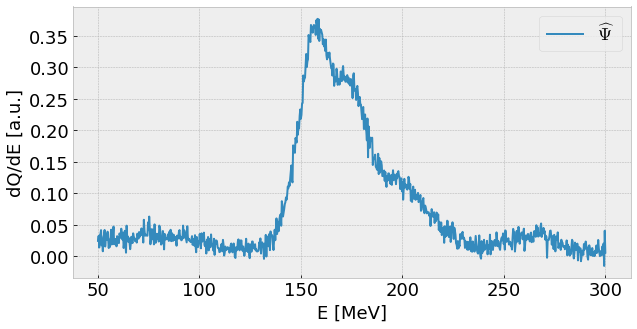}
    \caption{}
  \end{subfigure}
  \hfill
  \begin{subfigure}[t]{\textwidth}
    \includegraphics[width=12cm]{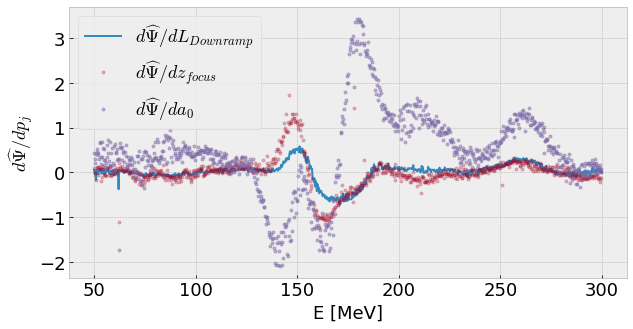}
    \caption{}
  \end{subfigure}
  \hfill
\caption{Figure (a) shows the predicted energy spectrum $\widehat{\Psi}$ at $a_0$=5, $z_\mathrm{focus}$=1e-4, $L_\mathrm{downramp}$=3.5e-5. Another advantage of iLWFA is that we are able to unveil the local contribution of certain input parameters $p_j$ with $p_0 = a_0$, $p_1=z_\mathrm{focus}$, $p_2=L_\mathrm{downramp}$ and $j\leq2$ to  $\widehat{\Psi}$. A significant and widespread contribution of e.g. $a_0$ implies that it might also be easier to reconstruct that parameter since the semantic region is large respectively (b). Additionally, this analysis might also stimulate further physics research on the non-linear contribution of simulation parameter to simulation outcome. }
\label{fig:attribution}
\vskip -0.05in
\end{figure}

The performance of our iLWFA model was assessed by randomly splitting our dataset into training and validation data. We found that the approximation of the forward pass yields a mean squared error of $MSE<0.007$ meaning that the energy spectra were recovered reasonable well (see e.g. Fig.~\ref{fig:energySpectra}). Furthermore, the similarity in shape of the reconstructed energy spectra with respect to groundtruth data was quantified by structured-similarity index  $SSIM\approx0.86$, i.e. the recovered energy spectra closely resemble the ground-truth data. The reconstruction of simulation parameters given an energy spectrum relates to solving an inverse problem. iLWFA allows us to solve this inverse problem by querying the \textit{invertible} neural network in reverse mode. A representative reconstruction of the posterior predictive distribtuion on validation data emphasizes the strengths of INNs for learning complex posterior distributions (see Fig.~\ref{fig:parameterReconstruction}). 
The median relative errors for reconstruction of simulation parameters from energy spectra is rather low (see table~\ref{tab:inverse}) meaning that the invertible network is learning features for joint forward simulation as well as parameter reconstruction. The good reconstruction performance on $a_0$ could be explained by strong contribution of that parameter to the reconstructed energy spectra (see Fig.~\ref{fig:attribution}). 

\begin{table}[t]
\caption{The relative error of the inverse pass are proportional to the contribution of each parameter to the energy spectrum.}
\label{tab:inverse}
\begin{center}
\begin{small}
\begin{sc}
\begin{tabular}{l c c}
\hline
relative error & training & validation \\
\hline
$a_0$ & $0.2\%$ & $0.3\%$ \\
$L_{\mathrm{downramp}}$ & $0.6\%$ & $3.7\%$ \\
$z_{\mathrm{focus}}$ & $2.5\%$  & $8.2\%$ \\
\hline
\end{tabular}
\end{sc}
\end{small}
\end{center}
\end{table}

\subsection{Direct physical application of the surrogate model }
Various beam parameters need to be tuned simultaneously to achieve a certain LWFA downramp injection for a targeted application. 
For examples narrow energy bandwidth $\Delta E $ around a high peak energy $E_\mathrm{peak}$ is required for operating Thomson scattering light sources \cite{Jochmann2013a,Kramer2018} or compact optical FELs \cite{Steiniger2017}. 
On the other hand, using these electron beams as driver in a compact plasma wakefield accelerator \cite{Kurz2020}, requires high-current beams, thus a high peak energy is needed but the requirements on $\Delta E$ are less strict.  

\begin{figure}[h]
\vskip 0.05in
\begin{center}

  \begin{subfigure}[t]{0.4\textwidth}
    \includegraphics[width=1.1\textwidth]{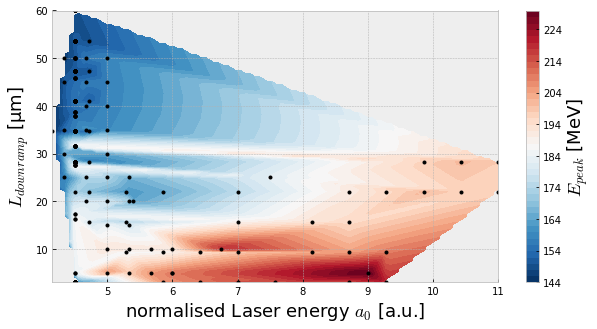}
    \caption{}
  \end{subfigure}
  \hfill
  \begin{subfigure}[t]{0.4\textwidth}
    \includegraphics[width=1.1\textwidth]{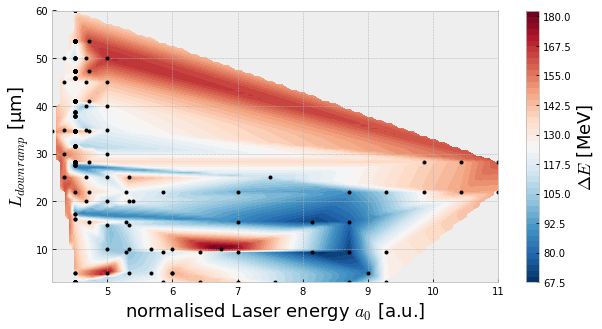}
    \caption{}
  \end{subfigure}

\caption{The surrogate model approximates the forward pass quite smoothly resulting in smooth transitions of our derived quantities peak energy $E_\mathrm{peak}$ with full-width half-maximum $\Delta E$. Hereby we are able to identify that highest peak energy can be reached with $a_0 \approx 9$ and a short downramp length $L_\mathrm{downramp}$ at a narrow energy bandwidth $\Delta E = \SI{82}{\mega \electronvolt}$.}
\label{fig:bunchparamter}
\end{center}
\vskip -0.05in
\end{figure} 


\section{Conclusions}
Laser wakefield acceleration is an established compact accelerator method promising significantly higher acceleration gradients than conventional particle accelerators. 
The numerical simulation of the involved complex physics requires jointly solving kinematic- and Maxwell's equation using particle-in-cell method.
Main contribution of this work is an invertible surrogate model, iLWFA, that is approximating the forward pass of a full LWFA simulation while providing all means for reconstruction of experimentally acquired quantities. 
Another benefit of ML-driven surrogate models is the differentiability of the model allowing us to infer physical knowledge from the incorporated non-linear mapping. 
An evaluation on large set of numerical LWFA simulations emphasizes the benefits of this approach but also outlines future work for making the reconstruction more robust.
\newpage
\bibliography{iclr2021_conference}
\bibliographystyle{iclr2021_conference}
\newpage
\section{Supplementary Material}

\end{document}